\documentclass[aps,prl,twocolumn]{revtex4}
\newcommand {\be}{\begin{equation}}
\newcommand {\ee}{\end{equation}}
\newcommand {\bey}{\begin{eqnarray}}
\newcommand {\eey}{\end{eqnarray}}
\newcommand {\mydoi}[2]{\href{http://dx.doi.org/#2}{#1}}
\usepackage{epsfig}
\usepackage{hyperref}

\begin{document}

\title{Epistemic view of quantum states and communication complexity of quantum channels}

\author{Alberto Montina}
\affiliation{Perimeter Institute for Theoretical Physics, 31 Caroline Street North, Waterloo, 
Ontario N2L 2Y5, Canada}

\date{\today}

\begin{abstract}
%The communication complexity of a quantum channel is the minimal amount of classical 
%communication required for simulating the channel. This quantity establishes 
%an upper bound for the communication saving provided by quantum channels over classical
%channels in a distributed computing scenario. 

The communication complexity of a quantum channel is the minimal amount of classical communication 
required for classically simulating a process of state preparation, transmission through the channel 
and subsequent measurement. It establishes a limit on the power of quantum communication in terms of 
classical resources. We show that classical simulations employing a finite amount of communication 
can be derived from a special class of hidden variable theories where quantum states represent
statistical knowledge about the classical state and not an element of reality. This special class has 
attracted strong interest very recently. The communication cost of each derived simulation is given by 
the mutual information between the quantum state and the classical state of the parent hidden variable 
theory. Finally, we find that the communication complexity for single qubits 
is smaller than $1.28$ bits. The previous known upper bound was $1.85$ bits.
\end{abstract}
\maketitle

\textit{Introduction.}---
As quantum systems can be simulated through classical resources, quantum theory is not 
intrinsically irreducible to an {\it ontological theory} picturing some well-defined and 
sharp reality through a classical language. The simplest way to realize this reduction 
is to interpret the quantum state as something real, such as a classical field. The 
definiteness of macroscopic reality would be guaranteed, for example, by some collapse 
mechanism, in the spirit of collapse theories, or by some auxiliary hidden variables, like 
in pilot wave theories. However, since the full information on the quantum state cannot be 
recovered by measurements on single systems, a more subtle classical theory would encode 
this information in the statistical behaviour of the classical state (hereafter called 
{\it ontic state}). Like a probability distribution, the quantum state would be asymptotically 
recovered from the ontic states of many identically prepared systems and would represent only
statistical knowledge about the actual state of single systems. These hypothetical theories 
are called $\psi$-epistemic. Conversely, theories where the quantum state information is contained 
in the ontic state of each realization are called $\psi$-ontic. The question whether 
the statistical encoding is actually possible has attracted growing interest in the recent 
years~\cite{hardy0,montina,spekkens,montina2,harri,montina3,montina4,bartlett,
montina5,pbr,lewis,colbeck,maxi,hardy}.
Indeed, the statistical role of the quantum state makes $\psi$-epistemic theories 
potentially less exposed to the principle of Occam's razor than $\psi$-ontic theories.
They could even be supported by the law of parsimony, as suggested in Ref.~\cite{montina}
and, more recently, in Refs.~\cite{montina4,montina5}. 

In this paper, we will show that $\psi$-epistemic theories have a pivotal role also in quantum 
communication and can determine an upper bound for the communication complexity of a 
quantum channel. Roughly speaking, we define the {\it communication complexity}, ${\cal C}_{min}$, 
of a quantum channel as the minimal amount of one-way classical communication required for 
simulating the channel through a classical protocol. The definition will be made more precise 
later. This quantity is important since it establishes a limit on the communication saving 
provided by quantum channels over classical channels in a distributed computing scenario. 
Indeed, it is known that, for some problems of distributed 
computing, quantum channels can provide huge communication saving over classical 
channels~\cite{buhrman}. However, a quantum channel cannot replace an amount
of classical communication greater than the communication complexity of that
channel. Thus, this latter quantity gives an upper bound for the power of
a quantum channel in terms of classical bits. 
At the present, it is not known if ${\cal C}_{min}$ is finite, apart from the case of single 
qubits~\cite{cerf,toner}. In Ref.~\cite{cerf}, it was shown that $2.19$ bits on average 
are sufficient for simulating the communication of a qubit. Toner and Bacon eventually 
improved this result by reporting a protocol requiring exactly $2$ bits of communication 
for each realization~\cite{toner}. In the case of many simulations performed in parallel,
the Toner-Bacon model gives an amortized communication cost equal to about $1.85$ bits.

A classical protocol that simulates quantum communication through a finite amount of classical 
communication (hereafter, more concisely, {\it finite communication protocol} or {\it FC protocol}) 
is clearly a kind of $\psi$-epistemic theory, since it is not possible to encode the full 
information on the quantum state in a finite number of classical bits. We will show that also 
the opposite is somehow true. More precisely, FC protocols can be derived from $\psi$-epistemic 
theories, provided that the mutual information, $I_m$, between the quantum state and the ontic 
state is finite. The communication cost, $\cal C$, is essentially 
given by $I_m$, that is, ${\cal C}\simeq I_m$. In the case of many simulations performed 
in parallel, the amortized asymptotic communication cost, say ${\cal C}^{par}$, turns out 
to be exactly equal to $I_m$.
As just FC protocols are known only for single qubits, at the present $\psi$-epistemic 
models with finite $I_m$ are known only for single qubits. One model 
was provided by Kochen and Specker in 1969~\cite{ks}. Using our result, we will show that 
the Kochen-Specker model implies that there is a FC protocol with 
${\cal C}^{par}=2-(2\log_e2)^{-1}\simeq1.28$ bits. This value
lowers previously known upper bounds for the communication complexity of single
qubits. It is quite surprising that this improved upper bound is derivable from a 
foundational result published about forty years ago and that this derivation went 
unnoticed for a long time. A protocol for simulating communication of single qubits by 
using exactly $2$ bits for each shot was derived from the Kochen-Specker model very
recently~\cite{montina7}. Our main purpose is to show that there is a 
relationship between $\psi$-epistemic theories with finite $I_m$ and FC
protocols.

The paper is organized as follows. After introducing the framework of an ontological
theory, we define the communication complexity ${\cal C}_{min}$ of quantum channels
and its generalized version, ${\cal C}^{par}_{min}$, for simulations performed in
parallel. Then, we present the main result concerning the relation between 
$\psi$-epistemic theories and FC protocols. As an 
illustration of this result, we derive the inequality ${\cal C}^{par}_{min}\le1.28$ bits
for single qubits from the Kochen-Specker model. 
Finally, the conclusions and perspectives are drawn.

\textit{Ontological theories.}---
Let us introduce the general framework of an ontological theory describing the 
preparation and the subsequent measurement of a quantum system.
In the ontological theory, a system is described by a set of variables, which we
denote by $x$. The value of $x$ represents the ontic state of the system. When the 
system is prepared in some quantum state $|\psi\rangle$, the ontic state $x$ is
set according to a probability distribution $\rho(x|\psi)$ that depends on 
$|\psi\rangle$. Thus, there is a mapping
\be\label{onto_distr}
|\psi\rangle\rightarrow \rho(x|\psi)
\ee
that associates each quantum state with a probability distribution on the ontological
space. A general measurement is described by a positive-operator valued measure (POVM),
which is defined by a set of positive semidefinite operators, 
$\{\hat E_1,\hat E_2,\dots\}\equiv{\cal M}$. Each operator $\hat E_i$ labels an 
event of the measurement $\cal M$. In the framework of an ontological theory, the 
probability of $\hat E_i$ is conditioned by the ontic state $x$. Thus, each measurement
$\cal M$ is associated with a probability distribution $P(\hat E_i|x,{\cal M})$,
\be
{\cal M}\rightarrow P(\hat E_i|x,{\cal M}).
\ee
Finally, the ontological theory is equivalent to quantum theory if the probability
of having $\hat E_i$ given the preparation $|\psi\rangle$ is equal to the quantum
probability, that is,
\be
\int dx P(\hat E_i|x,{\cal M}) \rho(x|\psi)=\langle\psi|\hat E_i|\psi\rangle.
\ee

In a $\psi$-ontic theory, the ontic state $x$ always contains the full information about 
the quantum state, that is, if $|\psi_1\rangle\ne|\psi_2\rangle$, then the distributions 
$\rho(x|\psi_1)$ and $\rho(x|\psi_2)$ are not overlapping. Conversely, in a $\psi$-epistemic 
theory, the ontic state does not contain this information, which is instead encoded only in 
the statistical behaviour of $x$, that is, in the distribution $\rho(x|\psi)$. 
In the class of $\psi$-epistemic theories, there is a subclass that is particularly 
relevant for the present discussion. In this subclass, the mean entropy of the probability
distribution $\rho(x|\psi)$ is finite, as well as the entropy of the maximally mixed
distribution $\int d\psi\rho(x|\psi)\rho(\psi)\equiv \rho(x)$, where $\rho(\psi)$ is a 
uniform distribution over the Hilbert space. In other words, the support of $\rho(x|\psi)$ and 
$\rho(x)$ have finite nonzero measure. We call the theories in this subclass {\it completely 
$\psi$-epistemic theories}. The Kochen-Specker model~\cite{ks} is an example of completely 
$\psi$-epistemic theory.

\textit{Classical simulation of quantum channels.}---
A quantum channel is a physical device, such as a wire, carrying information from a sender to 
a receiver. Mathematically, it corresponds to a map from a density operator to another density
operator. This map is completely positive and trace preserving.  Clearly, it is not 
possible to replace a quantum channel with a classical channel, unless entangled 
systems are shared between the sender and the receiver, like in quantum teleportation. Thus,
a classical simulation of a quantum channel simulates more properly a process where a party,
say Alice, prepares the input of a quantum channel in a quantum state $|\psi\rangle$ and
another party, say Bob, performs a subsequent measurement ${\cal M}$ 
on the quantum channel output. 
In this process, the quantum state corresponds operationally to a procedure described in classical 
terms. In other words, Alice has a classical description of the quantum state $|\psi\rangle$.
Without loss of generality, we assume that the quantum channel is a noiseless identity map,
since any noise and transformation can be transferred to the measurement stage by changing
the set of allowed measurements. 
%The set of prepared quantum states and measurements could have some possible constraint. 

A classical simulation of a quantum channel is as follows. Alice has a classical description of 
the quantum state $|\psi\rangle$ and she generates a variable $k$ with a 
probability $\rho(k|\chi,\psi)$ depending on the quantum state and a possible random variable, 
$\chi$, shared with Bob. The variable $\chi$ is generated according to the probability distribution
$\rho_s(\chi)$. Alice communicates the value of $k$ to Bob.
Finally, Bob generates an outcome $\hat E_i$ with a probability 
$P(\hat E_i|k,\chi,{\cal M})$. The protocol simulates exactly the quantum channel if the
probability of $\hat E_i$ given $|\psi\rangle$ is equal to the quantum probability,
that is, if
\be
\sum_k\int d\chi P(\hat E_i|k,\chi,{\cal M}) \rho(k|\chi,\psi)\rho_s(\chi)=\langle\psi|\hat E_i|\psi\rangle.
\ee
The communication cost $\cal C$ of a 
classical simulation can be defined in different ways. Here, we define $\cal C$ as the Shannon 
entropy of the distribution $\rho(k|\chi)\equiv\int d\psi\rho(k|\chi,\psi)\rho(\psi)$ averaged
over $\chi$,
\be
{\cal C}\equiv-\int dk\int d\chi\rho(k|\chi)\rho_s(\chi)\log_2\rho(k|\chi).
\ee
This definition makes sense if a probability distribution $\rho(\psi)$ of the quantum state 
$|\psi\rangle$ is given. A natural choice is to take $\rho(\psi)$ uniformly distributed.
If $N$ simulations are performed in parallel, it is possible to envisage a larger set of 
communication protocols, where the probability of generating $k$ can depend
on the full set of quantum states, say $|\psi_{i=1,2,\dots,N}\rangle$, prepared in 
each single simulation. In other words, the distribution $\rho(k|\chi,\psi)$ becomes
$\rho(k|\chi,\psi_1,\psi_2,\dots,\psi_{N})$. The asymptotic amortized communication cost, 
${\cal C}^{par}$, is the cost of the parallelized simulation divided by $N$ in the
limit of large $N$.
We define the {\it communication complexity} ${\cal C}_{min}$ of a quantum channel as the minimal 
amount of classical communication required by an exact classical one-shot simulation of the
quantum channel. Its asymptotic version for simulations performed in parallel, say 
${\cal C}^{par}_{min}$, has an obvious similar definition.

A classical protocol for simulating a quantum channel is essentially an ontological
theory, where the communicated index $k$ and the shared variable $\chi$ play the role of the
classical variable $x$. 
If the communication cost $\cal C$ of the classical protocol is finite, then the protocol
corresponds to a completely $\psi$-epistemic theory. In the following we will show that
also the opposite is true in some sense.

\textit{Communication cost from mutual information.}---
FC protocols can be derived from completely 
$\psi$-epistemic theories. To prove this, we use a result reported in Ref.~\cite{winter}
and its one-shot version~\cite{harsha}. Given two stochastic variables, say $x_1$ and $x_2$, 
with probability distribution $\rho(x_1,x_2)$, their mutual information,
$$
I_m=\sum_{x_1,x_2}\rho(x_1,x_2)\log_2\frac{\rho(x_1,x_2)}{\rho(x_1)\rho(x_2)},
$$
is the information that the two variables share. This interpretation of mutual information
suggests the following question. Suppose that the variable $x_1$ is generated by Alice
with marginal distribution $\rho(x_1)$. She sends Bob some amount of information, say 
$\cal C$, about $x_1$. Bob is required to set $x_2$ according to the joint 
distribution $\rho(x_1,x_2)$. What is the minimal value of $\cal C$ required for achieving
this goal? We expect that this value is given by the mutual information $I_m$. Indeed,
Winter proved in Ref.~\cite{winter}, that the amortized communication cost, in the asymptotic
limit of many simulations performed in parallel, is actually the mutual information,
that is,
\be
{\cal C}^{par}=I_m.
\ee
In general, the simulation protocol requires that Alice and Bob share some 
random variable $\chi$, which is uncorrelated with $x_1$.
A one-shot version of this result was recently reported in Ref.~\cite{harsha}.
For independently simulated realizations, we have that
\be\label{cost_one-shot}
I_m\le{\cal C}\le I_m+2\log_2(I_m+1)+2\log_2e.
\ee
Thus, the communication cost is the mutual information plus a possible
small additional cost that does not grow more than the logarithm of $I_m+1$.

\begin{figure}
\epsfig{figure=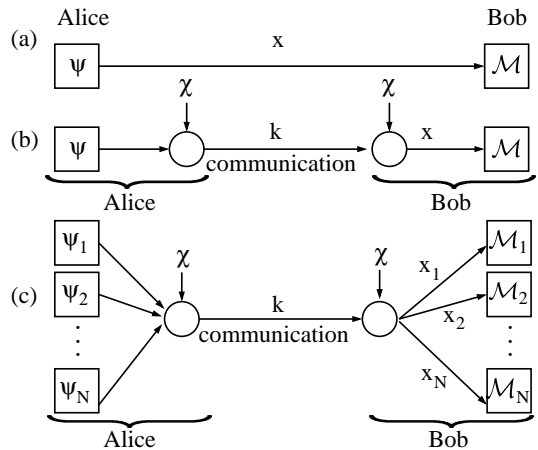,width=7.cm}
\caption{(a) Schematic representation of an ontological model. (b) The communication cost of
the model in (a) is reduced by demanding that Bob generates $x$. (c)
Multiple simulations performed in parallel.}
\label{fig1}
\end{figure}

These results have an immediate application to the problem of deriving FC protocols
from $\psi$-epistemic theories. An ontological theory is essentially a classical 
protocol for simulating quantum channels, where the ontic state $x$ is the communicated 
classical variable, as schematically illustrated in 
Fig.~\ref{fig1}. However, the communication cost can be, in general, infinite.
A strategy for making the communication of a $\psi$-epistemic theory finite and as small as
possible is as follows. Instead of communicating directly $x$, Alice can communicate 
an amount of information, encoded in an index $k$, that allows Bob to generate $x$ 
according to the probability distribution $\rho(x|\psi)$ (Fig.~\ref{fig1}b). By
Eq.~(\ref{cost_one-shot}), the minimal amount of required communication is
essentially equal to the mutual information between the quantum state $|\psi\rangle$
and the ontic state $x$, 
\be
{\cal C}\sim \int d\psi\int dx \rho(x,\psi)\log_2\frac{\rho(x,\psi)}{\rho(x)\rho(\psi)}\
\equiv I(X:\Psi).
\ee
If many simulations are performed in parallel (Fig.~\ref{fig1}c), Winter's result implies 
that there is a classical simulation such that the amortized asymptotic communication cost
is strictly equal to the mutual information,
\be\label{asym}
{\cal C}^{par}=I(X:\Psi).
\ee

It is clear that $I(X:\Psi)$ is infinite in $\psi$-ontic theories, since the ontic 
state shares with $|\psi\rangle$ the full infinite information contained in the quantum 
state. Conversely, completely $\psi$-epistemic theories have $I(X:\Psi)$ finite.

\textit{Application.}--- To illustrate the power of the above procedure,
we derive an upper bound for ${\cal C}_{min}^{par}$ from the Kochen-Specker model by
using Eq.~(\ref{asym}).
The Kochen-Specker model is an ontological model working for single qubits. The ontic state 
is given by a unit three-dimensional vector, $\vec x$. Let us represent a pure quantum
state through the unit Bloch vector, $\vec v$. Given the quantum state $\vec v$,
the probability distribution of $\vec x$ is
\be
\rho(\vec x|\vec v)=\pi^{-1}\vec v\cdot\vec x\theta(\vec v\cdot\vec x),
\ee
where $\theta$ is the Heaviside step function. Taking a uniform distribution for
$\vec v$, 
\be
\rho(\vec v)=(4\pi)^{-1},
\ee
the mutual information $I_m$ between $\vec x$ and $\vec v$ is 
\be
%\begin{array}{c}
I_m=
%\frac{1}{\pi}\int d^2v\int d^2x \theta(\vec v\cdot\vec x)
%\vec v\cdot\vec x\log_2 (4 \vec v\cdot\vec x)=  
%\vspace{1.5mm}  \\
%2\int_0^{\pi/2}d\eta \sin\eta\cos\eta\log_2(4\cos\eta)=
%\vspace{1.5mm}\\
2-(2\log_e2)^{-1}\simeq 1.28 \text{ bits}.
%\end{array}
\ee
This implies that
\be
{\cal C}_{min}^{par}\le 1.28 \text{ bits}.
\ee
To the best of our knowledge, this upper bound improves previously known results.
The amortized asymptotic cost of the model in Ref.~\cite{cerf} is about $2.19$ bits,
whereas the model of Ref.~\cite{toner} requires about $1.85$ bits of communication.

The Kochen-Specker model works only for single qubits. Very recently, a $\psi$-epistemic
model for higher dimensional systems was introduced in Ref.~\cite{lewis}. Unfortunately, 
in this model, there is a finite probability that the ontic state contains the full
information about the quantum state, this makes $I(X:\Psi)$ infinite. Thus, the model 
is not completely $\psi$-epistemic and its associated communication cost is infinite.

\textit{Conclusion.}---
In this paper, we have shown that there is relationship between $\psi$-epistemic theories
and classical protocols that simulate quantum channels through a finite amount $\cal C$ 
of communication (FC protocols). Indeed, FC protocols are particular kinds
of $\psi$-epistemic theories. We have shown that also the opposite is somehow true.
More precisely, given a completely $\psi$-epistemic model, we have described a procedure for 
turning the model into a FC protocol with $\cal C$ 
essentially equal to the mutual information between the quantum state and the ontic 
state.
If many simulations are performed in parallel, then there is a procedure for turning
the $\psi$-epistemic model into a global simulation of all the quantum channels
with amortized asymptotic cost ${\cal C}^{par}$ exactly equal to the mutual 
information. Using this result, we have shown that the Kochen-Specker model
can be turned into a classical protocol of communication with 
${\cal C}^{par}\simeq 1.28$ bits.

The main motivation of this paper is to show that $\psi$-epistemic theories, which
are attracting increasing interest in quantum foundation, have a relevant role also 
in quantum communication. The acknowledgement of this relationship can provide a useful
cross-fertilization between two related fields. Indeed, some recent general results
about $\psi$-epistemic theories have a direct consequence in quantum communication.
For example, the result in Ref.~\cite{montina2} implies that the number of continuous 
shared variables in a FC protocol cannot be 
smaller than $2N_h-2$, $N_h$ being the Hilbert space dimension, provided that the 
shared noise and the communicated variable satisfy a suitable transformation rule. 
Also the recent result reported in Ref.~\cite{pbr} has a consequence in quantum 
communication. 
%The ontological excess baggage theorem~\cite{hardy2} is a further example 
%of this overlap between quantum foundation and quantum information, as
%it is equivalent to assert that a classical protocol without shared randomness 
%cannot simulate a quantum channel with a communication cost being finite in the worse 
%case, proved in Ref.~\cite{massar}. 
Furthermore,
the procedure of reducing $\psi$-epistemic theories to FC protocols can make easier to 
find FC protocols, as the class of completely $\psi$-epistemic theories is larger than
the class of FC protocols. Indeed, in a subsequent work, we will use the procedure
discussed here for proving that the communication cost of classically simulating the 
communication of $n$ qubits is independent of $n$, provided that a bounded error is 
allowed and only rank-$1$ projective measurements are performed~\cite{future}.

{\it Acknowledgements.} 
The Author acknowledges useful discussions with Jonathan Barrett and Iman Marvian.
Research at Perimeter Institute for Theoretical Physics is
supported in part by the Government of Canada through NSERC
and by the Province of Ontario through MRI.

\bibliography{biblio.bib}

\end{document}